\documentclass[english,aps,prl,notitlepage,twocolumn]{revtex4-1}
\usepackage[T1]{fontenc}
\usepackage{color}
\usepackage{amsmath}
\usepackage{amssymb}
\usepackage{graphicx}
\usepackage{bm}
\usepackage[normalem]{ulem}

\usepackage{babel}
\global\long\def\ket#1{\left| #1\right\rangle }

\global\long\def\bra#1{\left\langle #1\right|}

\global\long\def\braket#1#2{\left\langle #1| #2\right\rangle }

\global\long\def\mk{\bm{k}}

\global\long\def\mr{\bm{r}}

\usepackage{lipsum}

\makeatletter

\usepackage{amsmath}
\usepackage{endnotes}
\makeatother

\usepackage{babel}
\begin{document}
\global\long\def\ket#1{\left| #1\right\rangle }

\global\long\def\bra#1{\left\langle #1\right|}

\global\long\def\braket#1#2{\left\langle #1| #2\right\rangle }

\global\long\def\mk{\bm{k}}

\global\long\def\mr{\bm{r}}

\title{{Floquet superradiance lattices in thermal atoms}}

\author{Xingqi Xu$^{1,*}$, Jiefei Wang$^{1,2*}$, Jianhao Dai$^{1}$, Ruosong Mao$^{1}$, Han Cai$^{2,1\dagger}$, Shi-Yao Zhu$^{1,3}$, and Da-Wei Wang$^{1,3,4\ddagger}$}

\affiliation{$^1$Interdisciplinary Center for Quantum Information, State Key Laboratory of Modern Optical Instrumentation, and Zhejiang Province Key Laboratory of Quantum Technology and Device, Department of Physics, Zhejiang University, Hangzhou 310027, Zhejiang Province, China;\\
$^2$College of Optical Science and Engineering, Zhejiang University, Hangzhou, 310027, China;\\
$^3$Hefei National Laboratory, Hefei 230088, China;\\
$^4$CAS Center for Excellence in Topological Quantum Computation, University of Chinese Academy of Sciences, Beijing 100190, China}

\begin{abstract}
Floquet modulation has been widely used in optical lattices for coherent control of quantum gases, in particular for synthesizing artificial gauge fields and simulating topological matters. However, such modulation induces heating which can overwhelm the signal of quantum dynamics in ultracold atoms. Here we report that the thermal motion, instead of being a noise source, provides a new control knob in Floquet-modulated superradiance lattices, which are momentum-space tight-binding lattices of collectively excited states of atoms. The Doppler shifts combined with Floquet modulation provide effective forces along arbitrary directions in a lattice in frequency and momentum dimensions. Dynamic localization, dynamic delocalization and chiral edge currents can be simultaneously observed from a single transport spectrum of superradiance lattices in thermal atoms. Our work paves a way for simulating Floquet topological matters in room-temperature atoms and facilitates their applications in photonic devices.
\end{abstract}

\maketitle

{Collective interaction between atoms and light plays an important role in quantum optics \cite{Gross1982,Inouye1999}. A most famous example is the Dicke superradiance featured by enhanced decay rates of collectively excited atoms \cite{Dicke1954}, in particular for atoms confined in a small volume compared with the light wavelength. In the opposite limit, i.e., when the size of the atomic cloud is much larger than the light wavelength, Scully {\itshape et al.} \cite{Scully2006,Scully2009} found that the light momentum can be stored in atoms prepared in the so-called timed Dicke states (TDSs) \cite{He2020,Araujo2016,Roof2016}. The stored momentum can be released in the directional superradiant emission \cite{Scully2006}. More interestingly, such TDSs with different momenta can be coupled by multiple lasers to form momentum-space superradiance lattices (SLs) \cite{Wang2015,Wang2015optica}, which provide a new platform for quantum simulation. 

In contrast to real-space optical lattices that require ultracold temperature, SLs can be implemented in both cold atoms \cite{Chen2018,Wang2020npj,Mi2021} and room-temperature atoms \cite{Cai2019,He2021,Mao2022}, since TDSs are robust to the center-of-mass motion of atoms \cite{Celardo2014,Damanet2016,Bromley2016,Weiss2019}.}~Such a robustness can be understood from a distinct feature of the {momentum-space} SL, i.e., its Brillouin zone (BZ) is in real space.
Here the positions of atoms play the same role of the crystal momenta of electrons in solids. Atoms moving through the real-space BZs follow the same dynamics of electrons in a DC electric field. Therefore, the atomic motion, instead of being a source of noise, can simulate an effective electric {field} in momentum space, which has been used to measure the geometric Zak phase \cite{Mao2022,Zak1989,Xiao2010}.

Here we show that SLs can also be used to study the rich physics in Floquet (temporally periodic) systems \cite{Eckardt2017,Oka2019,Rudner2020,Weitenberg2021,Torre2021}, where the thermal motion induced DC {field} significantly enriches the tunability. Being a widely-used technique in quantum engineering \cite{Oka2009,Lindner2011,Rechtsman2013,Jotzu2014,Rudner2013,Mukherjee2017,Cheng2019,Jiang2011,Titum2016,Kolodrubetz2018,Nathan2021,Li2022}, Floquet  driving \cite{Floquet,Shirley1965,Leon2013,Goldman2014} can extend the 1D SLs into (1+1)D (momentum-frequency dimensional) lattices, allowing us to simulate higher-dimensional lattice dynamics \cite{Martin2017,Boyers2020,Long2021,Malz2021}.  In the (1+1)D SLs, there is a constant effective electric {field} in the frequency dimension, determined by the Floquet modulation frequency \cite{Dutt2021}. 
On the other hand, the atomic velocity provides a DC {field} along the momentum dimension with a strength proportional to the Doppler shift. By resolving the direction of the total {field} with a spectroscopic method, we observe dynamic localization, dynamic delocalization and chiral edge currents \cite{Dunlap1986,Holthaus1992,Madison1998,Longhi2006,Lignier2007,Sias2008,Szameit2009,Szameit2010,Haller2010,Ma2011,Chen2011,Yuan2015,Mancini2015,Stuhl2015,Libi2016,Atala2014,An2017,Wang2020} in SLs of thermal atoms.  


\begin{figure}[!htb]
\centering
\includegraphics[width=0.95\columnwidth]{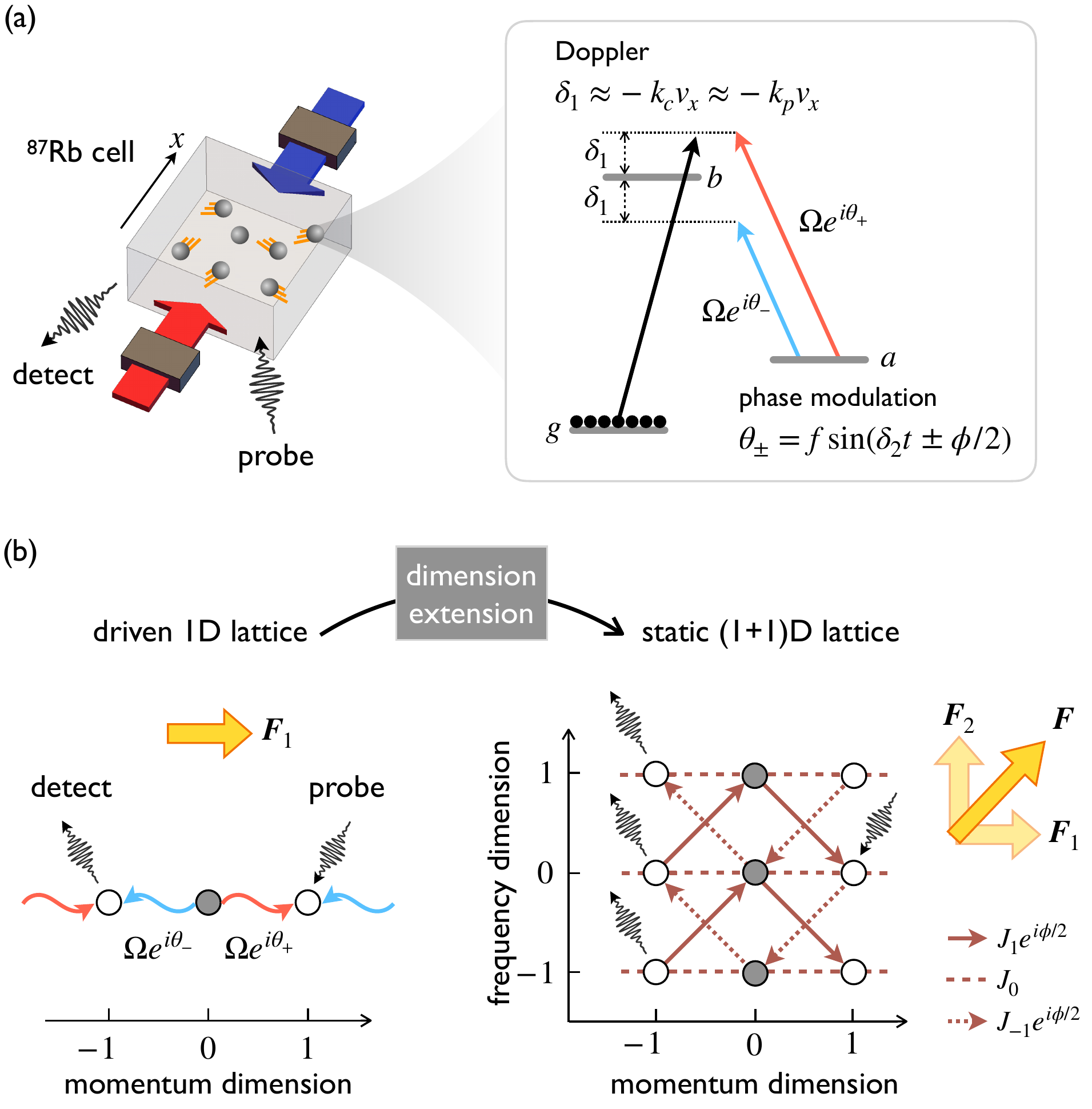}
\caption{(a) Schematic configuration of the experimental setup. The coupling fields are phased-modulated by two EOMs at a RF frequency $\delta_2=80$ MHz. Insets: the atomic levels in the reference frame of the atoms with velocity $v_x$. The coupling fields have opposite Doppler shifts $\pm\delta_1$ and periodically modulated phases $\theta_\pm(t)$. (b) Mapping between the dynamically modulated 1D SL and the (1+1)D Floquet SL, {with white and gray circles labeling the $b$- and $a$-sublattice sites, respectively.} Left: the driven 1D momentum-space lattices with phase-modulated hopping strengths. The yellow arrow $\bm{F}_1$ indicates the direction of atomic motion induced DC electric force. Right: the equivalent (1+1)D frequency-momentum lattices. The  yellow arrow $\bm{F}$ indicates the total DC electric force with components $\bm{F}_{1}$ and $\bm{F}_{2}$ along the momentum and frequency dimensions. {The probed and detected sites are marked schematically.}}
\label{fig1}%
\end{figure}

\textit{Experimental setup and theoretical model.} 
The experiment is implemented in a vapor cell of natural abundance rubidium atoms in a standing-wave-coupled electromagnetically induced transparency (EIT) configuration (see Fig.~\ref{fig1}(a)).  A weak probe field couples the ground state $|g\rangle\equiv|5^2S_{1/2},F=1\rangle$ to an excited state $|b\rangle\equiv|5^2P_{1/2},F=2\rangle$ in $^{87}$Rb D1 line. The excited state is resonantly coupled to a metastable state $|a\rangle\equiv|5^2S_{1/2},F=2\rangle$ by two strong laser fields. {The probe field pumps the atoms to the TDS,
\begin{equation}
|b_{\bm{k}_p}\rangle = \frac{1}{\sqrt{N}}\sum_{j=1}^{N}e^{i\bm{k}_p\cdot\bm{r}_j}|g_1...b_j...g_N\rangle,
\end{equation}
where $\bm{k}_p$ is the wavevector of the probe field, $\bm{r}_j$ is the position of the $j$th atom.~The standing-wave coupling fields couple $|b_{\bm{k}_p}\rangle$ to other TDSs with discrete momenta to form SLs \cite{Wang2015}, in which the directional superradiant emission from the TDS $|b_{-\bm{k}_p}\rangle$ is collected along $-\bm{k}_p$ direction \cite{Scully2006,Scully2009,Roof2016,Araujo2016,He2020}, such that the lattice transport from $|b_{\bm{k}_p}\rangle$ to $|b_{-\bm{k}_p}\rangle$ can be investigated. We can tune the probe frequency to measure the transport at different energies, similar to tuning the Fermi energy of electrons in solids.}


{In our experiment on 1D SLs, the probe field propagates in $+x$ direction with a wave vector $k_p$, parallel to the two coupling laser fields with wave vectors $\pm k_c$ and $k_p\approx k_c$}. The phases of the coupling fields are modulated by two electric-optic modulators (EOMs) independently, $\theta_\pm(t) = f\sin(\delta_2t\pm\phi/2)$, where $f$  and $\delta_2$ are the modulation depth and frequency, and $\phi$ controls the phase delay between the two radio-frequency (RF) driving fields. The interaction between the atoms and the coupling fields is described by a Floquet SL (see Fig.~\ref{fig1}(b)). The EOMs introduce periodic modulation of the coupling strengths in the SL Hamiltonian (we set $\hbar=1$ and see details in \cite{SM}), 
\begin{equation}
H_d= \sum_{j} \Omega a^{\dagger}_{2j}[e^{-i\theta_+(t)} b^{\ }_{2j+1}+e^{-i\theta_-(t)} b^{\ }_{2j-1}]+H.c.,
\end{equation}
where $\Omega$ is the Rabi frequency of the coupling fields, $a_{2j}^\dagger$ and $b_{2j+1}^\dagger$ are creation operators in the $j$th unit cell of the SL, $a_{2j}^\dagger\equiv 1/\sqrt{N} \sum_l e^{2ijk_cx_l}|a_l\rangle \langle g_l|$ and $b_{2j+1}^\dagger\equiv 1/\sqrt{N} \sum_l e^{i(2j+1)k_cx_l}|b_l\rangle \langle g_l|$ with $x_l$ being the position of the $l$th atom and $N$ the total number of atoms. They create TDSs from the ground state, e.g., $a_{2j}^\dagger |g_1, g_2,...,g_N\rangle=1/\sqrt{N}\sum_l \exp(2ijk_cx_l)|g_1...a_l...g_N\rangle$, which carries a momentum $2j k_c$. 



Using the Floquet theorem for $H_d$, we obtain an effective (1+1)D Hamiltonian $V_2+H_s$ \cite{SM}, including a  {potential energy in the frequency dimension \cite{Yu2021},}
\begin{equation}
V_2=-\delta_2\sum_{j,m}m(a^{\dagger}_{2j,m}a^{\ }_{2j,m}+b^{\dagger}_{2j+1,m}b^{\ }_{2j+1,m}),
\end{equation} 
and the static lattice Hamiltonian $H_s=\sum_nH_s^{[n]}$ with
\begin{equation}
\begin{split}
H_s^{[n]} = \sum_{j,m} &\Omega e^{in\phi/2} a^{\dagger}_{2j,m}[J_{-n}(f)b_{2j+1,m+n}\\
&+J_{n}(f)b_{2j-1,m-n}]+ H.c.,
\end{split}
\label{eq:H_l}
\end{equation}
{which couples lattice sites along the crystal direction $\hat{e}_1+n\hat{e}_2$ (denoted by Miller index $[1,n]$) with $\hat{e}_1$ and $\hat{e}_2$ being the unit vectors along the momentum and frequency dimensions.} Here $J_n$ is the $n$th order Bessel function and $d_{j,m}^\dagger$ ($d=a,b$) is the creation operator for {$|d_{j,m}\rangle\equiv e^{-im\delta_2 t}|d_{j}\rangle$ \cite{Leon2013, SM}}. We can understand $|d_{j,m}\rangle$ as a replica of $|d_j\rangle$ that carries $-m\delta_2$ energy from the periodic driving \cite{Gao2021}.  The $n$th order hopping coefficient carries a phase $\pm n\phi/2$, which can synthesize a gauge field in the lattice (see Fig.~\ref{fig1} (b)). The linear potential along the frequency dimension $V_2$ can be characterized by a force $\bm{F}_2=\delta_2\hat{e}_2$. Throughout the experiment we set $\delta_2=80$ MHz and $\Omega=25$ MHz to guarantee that the driving frequency is much larger than hopping strengths. 

For moving atoms, {the photons in the two coupling fields along $\pm x$ directions have extra energies due to the Doppler shifts $\pm\delta_1$ with $\delta_1=-k_cv_x$ and $v_x$ being the atomic velocity in $x$ direction.} Since the TDSs in the SL are created by absorbing and emitting photons in the coupling fields, the SL acquires a linear potential along the momentum dimension (in the lab reference frame), 
\begin{equation}
V_1=-\delta_1\sum_{j,m} [2j a_{2j,m}^\dagger a_{2j,m}+(2j+1) b_{2j+1,m}^\dagger b_{2j+1,m}],
\end{equation}
which has the effect of a force $\bm{F}_1=\delta_1 \hat{e}_1$. Consequently, the force associated to the total linear potential $V_1+V_2$ is $\bm{F}=\bm{F}_1+\bm{F}_2=\delta_1\hat{e}_1+\delta_2\hat{e}_2$. {In our experiment, the hopping of excitations along the force direction is inhibited because the potential difference is much larger than the coupling strengths.}

\begin{figure}[!t]
\centering
\includegraphics[width=0.8\columnwidth]{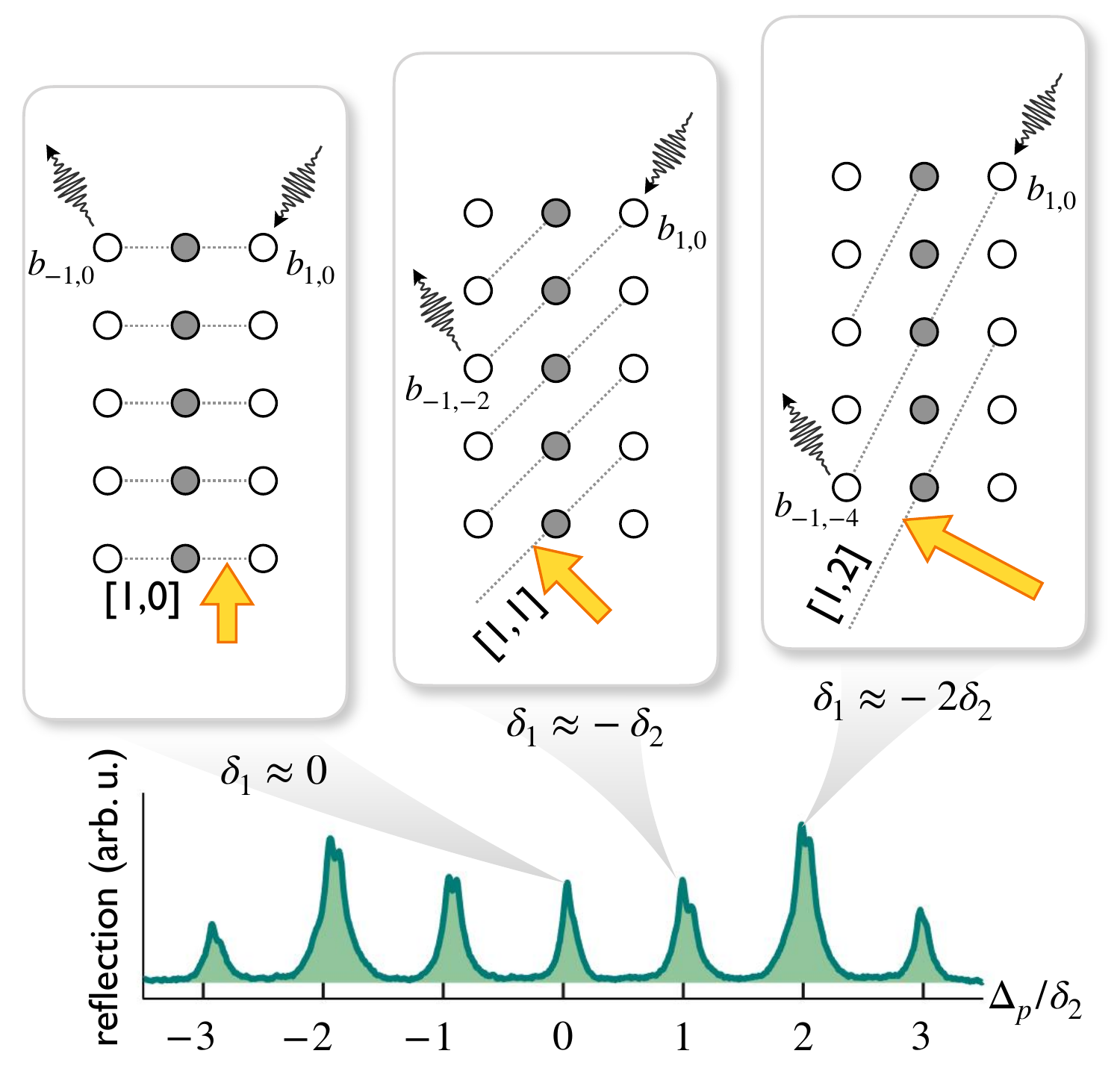}
\caption{Correspondence between peaks at $\Delta_p=n \delta_2$ in the reflection spectrum and the dynamics along the crystal direction $[1,n]$ (dashed lines). The $n$th peak is contributed by the atoms in the velocity group with a Doppler shift $\delta_1=-n\delta_2$. The effective force $\bm{F}=-\Delta_p\hat{e}_1+\delta_2\hat{e}_2$ makes the dynamics of (1+1)D lattices dominated by transitions along the crystal direction $[1,n]$. The reflection signal shows the transport from the probed state $|b_{1,0}\rangle$  to the detected site $|b_{-1,-2n}\rangle$. The phase delay $\phi=\pi$ and modulation depth $f=3$.}
\label{fig2}%
\end{figure}

\textit{Lattice transport measurement.}
We can tune the probe frequency to measure the transport in SLs with different values of $\bm{F}$. The interaction Hamiltonian between the probe field and the atoms is $H_p=\sqrt{N}\Omega_p e^{-i\Delta_p t}b^\dagger_{1,0}+ H.c.$, where $\Delta_p=\nu_p-\omega_{bg}$ with $\nu_p$ being the probe field frequency and $\omega_{bg}$ being the atomic transition frequency in the lab reference frame. The probe field couples the ground state to a TDS $|b_{1,0}\rangle$, which has a potential energy $-\delta_1$ proportional to the atomic velocity, such that we need to detune the probe field $\Delta_p\approx-\delta_1$ to excite the TDSs of atoms with the corresponding velocities. Since $\bm{F}_1$ is also proportional to the atomic velocity, we can measure lattice dynamics with different $\bm{F}_1$ at the corresponding probe detunings. In another word, with the probe detuning $\Delta_p$, we measure the transport from $|b_{1,0}\rangle$ to $|b_{-1,m}\rangle$ in an effective force $\bm{F}=-\Delta_p\hat{e}_1+\delta_2\hat{e}_2$. Such transport is experimentally measured from the {superradiant directional emission of $|b_{-1,m}\rangle$ along the $-x$ direction, which satisfies the phase-matching condition $-2k_c+k_p\approx -k_p$ \cite{SM}}. Therefore, we can simultaneously measure the dynamics with different effective forces in a single transport spectrum by detuning the probe field frequency. {The coherent probe field may also create more than one excitations which, however, have the same dynamic response as the TDSs due to the bosonic nature of atomic excitations in a weak probe field \cite{Wang2015,Cai2019}.}

\begin{figure}[!t]
\centering
\includegraphics[width=1\columnwidth]{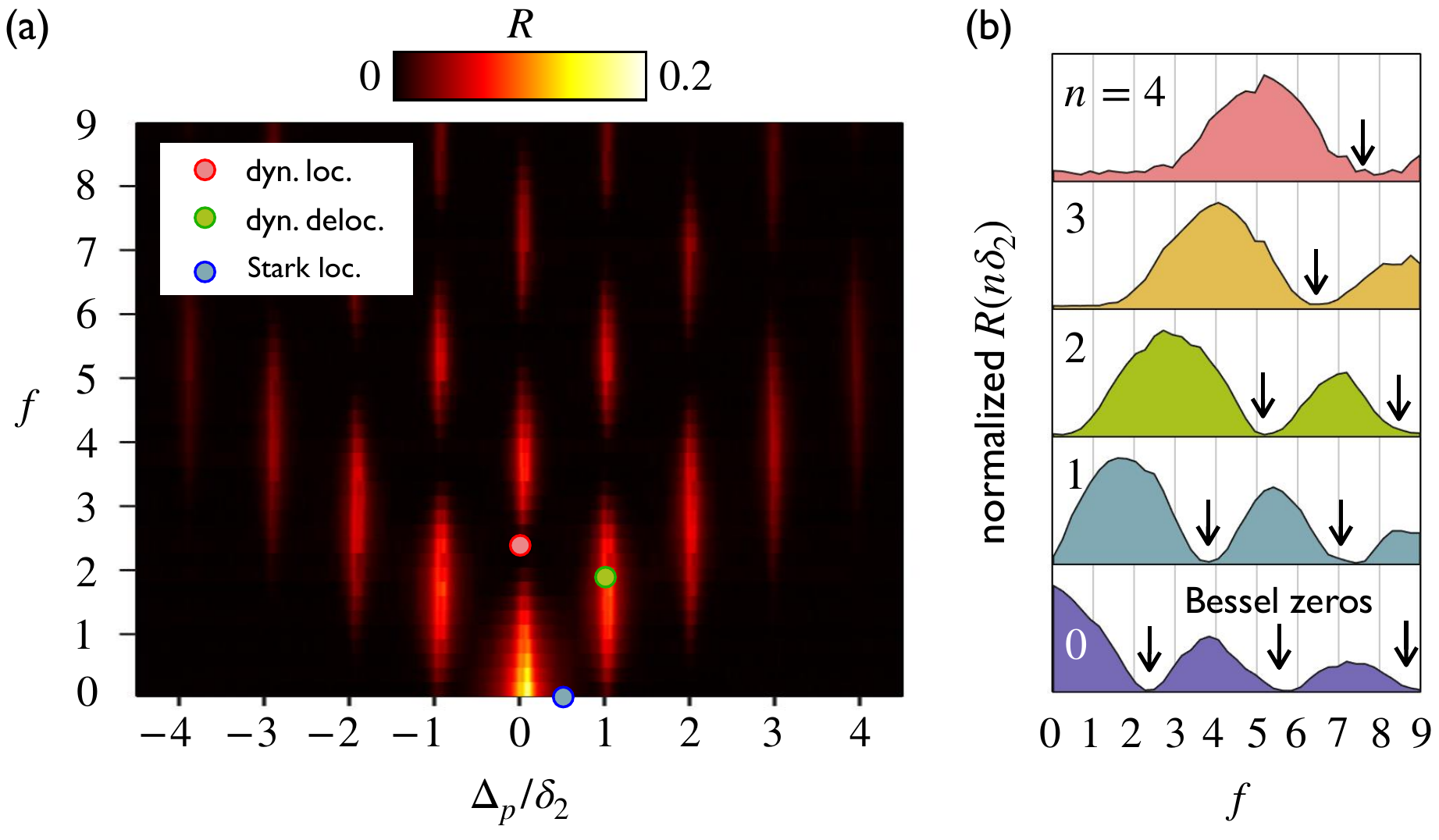}
\caption{Dynamic localization and delocalization. (a) Experimental reflection spectra as a function of the probe detuning $\Delta_p$ and the modulation depth $f$. We denote the points for Stark localization, dynamic localization, and dynamic delocalization with dots in blue, red and green colors. (b) The relfections at $\Delta_p=n\delta_2$ characterizing transport along crystal directions $[1,n]$ ($n=0,1,2,3,4$) are measured as functions of $f$, where the transport is completely inhibited at the zero points (marked by black arrow) of $J_n(f)$. We set the phase delay $\phi=\pi$. }
\label{fig3}%
\end{figure}

We show a typical spectrum for the transport from $|b_{1,0}\rangle$ to $|b_{-1,m}\rangle$ in Fig.~\ref{fig2}. Due to the Stark localization, the transport is only significant when the two states have the same potential energy, i.e., when the force points in a direction perpendicular to a crystal direction of the (1+1)D SL. As a result, the reflection is featured by discrete peaks at $\Delta_p=n\delta_2$. The reflection peak at $\Delta_p=n\delta_2$ is contributed by the transport from $|b_1,0\rangle$ to $|b_{-1,-2n}\rangle$ along the crystal direction $[1,n]$ (dotted lines in the insets), perpendicular to $\bm{F}$. From the reflection spectra we are able to investigate the dynamic localization and delocalization by simultaneously tuning the modulation depths of the two EOMs while keep their phase delay $\phi=\pi$ fixed. We can also tune $\phi$ to introduce effective magnetic fluxes in the (1+1)D SLs to investigate the Floquet chiral edge currents.

\begin{figure*}[!ht]
\centering
\includegraphics[width=1.4\columnwidth]{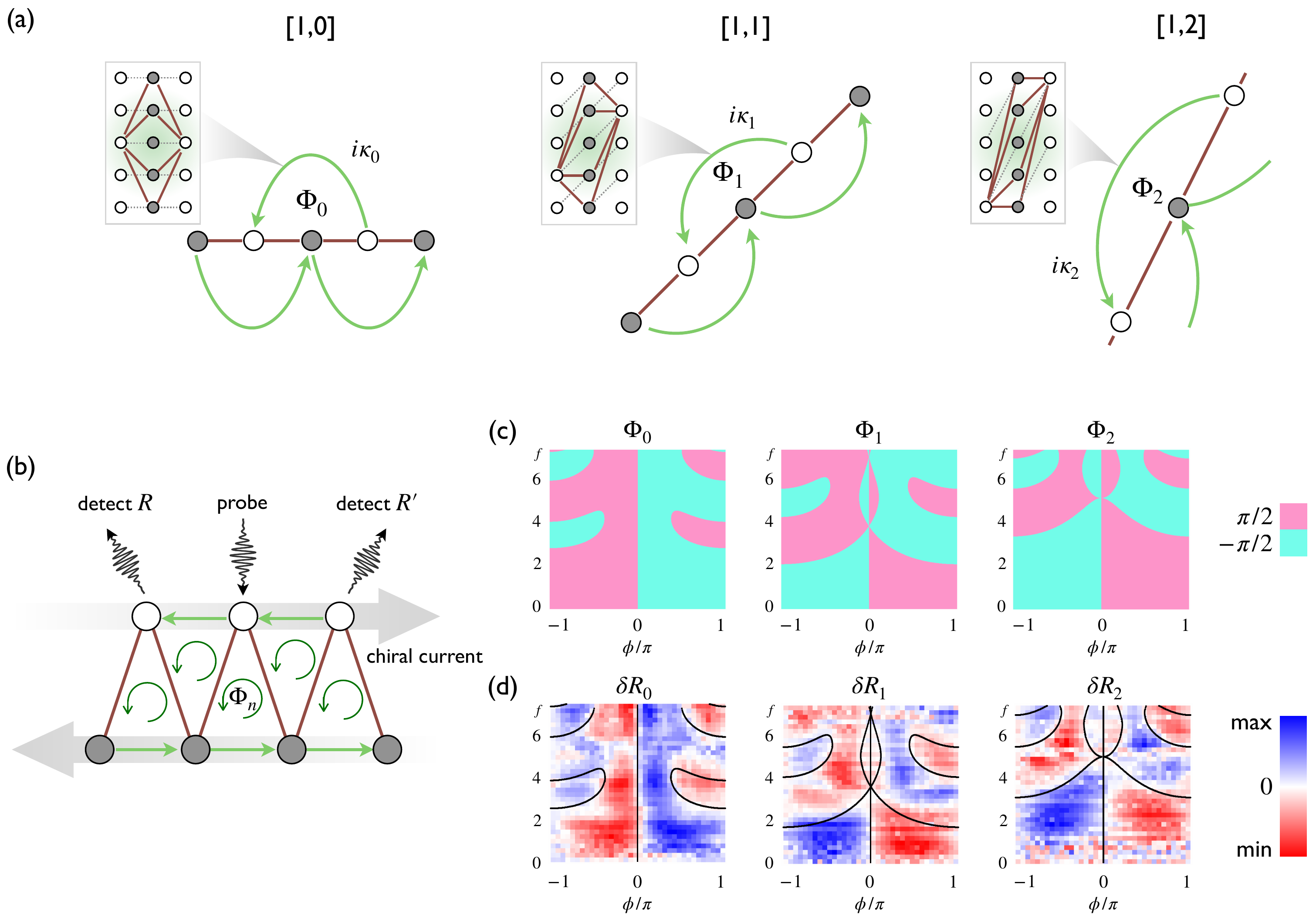}
\caption{Floquet chiral currents. (a) The schematics of the effective Hamiltonian along  a composite dimension of momentum and frequency. The NNN hoppings (green lines with arrows indicating the transition directions attached with a phase $\pi/2$) along crystal direction $[1,n]$ ($n=0,1,2$) are induced by all possible second-order transition paths (insets). (b) The zigzag chiral ladder along $[1,n]$ (replotted from the reduced 1D lattices in (a)) with an effective magnetic flux $\Phi_n$ in a unit loop. The two sublattices are the two edges which host chiral edge currents {with directions (along thick arrows) determined by $\Phi_n$. The green arrows denote the transition directions attached with a phase $\pi/2$.} (c) Phase diagram of $\Phi_n$ as functions of $\phi$ and $f$. (d) The phase diagram of chiral currents measured by $\delta R_n\equiv R(n\delta_1)-R'(-n\delta_1)$, which are consistent with the corresponding phase diagram of $\Phi_n$ with phase boundaries indicated by solid lines.}
\label{fig4}%
\end{figure*}

\textit{Dynamic localization and delocalization.}
We measure the reflection spectra as a function of the probe detuning $\Delta_p$ and the modulation depth $f$ (see Fig.~\ref{fig3}(a)). The transport along the crystal direction $[1,n]$ is demonstrated near the detuning $\Delta_p=n\delta_2$ and governed by the interaction Hamiltonian $H_s^{[n]}$ in Eq.~(\ref{eq:H_l}). Remarkably, the transport is totally inhibited when the modulation depth $f$ reaches to the zeros of the corresponding Bessel functions $J_n(f)$ (marked by black arrows in Fig.~\ref{fig3}(b)), where the effective hopping strength is zero in $H_s^{[n]}$. In particular, at the detuning $\Delta_p=0$, the reflectivity vanishes at $f=2.4$, $5.5$, and $8.7$, which are the first three zeros of $J_0(f)$, characterizing the dynamic localization \cite{Dunlap1986,Holthaus1992}. For parameters in our experiment, localization at zeros of  $J_n(f)$ can be clearly identified up to $|n|=4$, a significant improvement compared with other platforms \cite{Sias2008,Szameit2009}. The order $n$ is limited by the width of the Doppler broadening, which has a FWHM of $500$ MHz at the experiment temperature $60^\circ$C.

The measured transport can  be understood as the interplay between the lattice dynamics in AC and DC  {field} \cite{Dunlap1986,Holthaus1992,Madison1998,Longhi2006,Lignier2007,Sias2008,Szameit2010,Haller2010,Ma2011,Chen2011,Yuan2015} in a 1D SL. The oscillating phases $\theta_\pm(t)$ introduce a uniform AC  {field} $(\partial\theta_-/\partial t-\partial\theta_+/\partial t)/2 = f\delta_2\sin(\delta_2t)$ \cite{Leon2013}. We observe three important effects in the presence of both AC and DC {field} as being marked in  Fig.~\ref{fig3}(a). Without AC and DC {fields}, i.e., for $\delta_1=f=0$ near $\Delta_p=0$, the 1D SL transport signal is strong from $|b_1\rangle$ to $|b_{-1}\rangle$. When we keep $f=0$ and detuning $\Delta_p$ away from zero, the signal disappears due to Stark localization in a DC {field} (see the blue point at $\Delta_p=\delta_2/2$ and $f=0$). When we keep $\Delta_p=0$ and increase $f$, an effective AC {field} leads to the dynamic localization at $f=2.4$ \cite{Dunlap1986,Holthaus1992,Madison1998,Longhi2006,Lignier2007,Szameit2010}. With both AC and DC {fields}, the transport is recovered when the modulation frequency of the AC {field} is on resonance with the Bloch frequency of the DC {field} (see the green dot for $\Delta_p=\delta_2$), which is called dynamic delocalization or photon-assisted tunnelling \cite{Sias2008,Haller2010,Ma2011,Chen2011,Yuan2015}.

\textit{Phase diagram of chiral currents.}
{Chiral edge currents have been observed in zigzag SLs \cite{Cai2019}. Floquet engineering can bring new control knobs to induce such unidirectional transport \cite{Mancini2015,Stuhl2015,Libi2016,Atala2014,An2017,Wang2020}}.  Near $\Delta_p=n\delta_2$, the Hamiltonian $H_s^{[n]}$ introduces equipotential transitions while $H_s^{[m]}$'s with $m\neq n$ provide second-order transitions via intermediate states (see green lines in Fig.~\ref{fig4}(a) and derivation in \cite{SM}),
\begin{equation}
\begin{aligned}
K^{[n]}&=\sum_{j,m} i\kappa_n(b^\dagger_{2j-1,m-n}b^{\ }_{2j+1,m+n}\\
&-a^\dagger_{2j,m}a^{\ }_{2j+2,m-2n})+H.c.,
\end{aligned}
\label{eq:k}
\end{equation}
with  $\kappa_n=-\sum^{\infty}_{p=1} {2\sin(p\phi)}\Omega^2 J_{p+n}(f)J_{p-n}(f)/{p \delta_2}$ \cite{SM}. When $\phi=\pi$, we obtain $\kappa_n=0$ for dynamic localization and delocalization. In general, for $\phi\neq 0$ and $\pi$, {$H_s^{[n]}$ and $K^{[n]}$ induce the nearest-neighbor and next-nearest-neighbor (NNN) hoppings  (see red lines and green arrows in Fig.~\ref{fig4}(a) and (b)). The phase factors in $K^{[n]}$ bring a magnetic flux in a unit closed loop (circular arrows in Fig.~\ref{fig4}(b)) \cite{Cai2019,Hugel2014}},
\begin{equation}
\Phi_n=\frac{\pi}{2}(-1)^n\text{sgn}[\kappa_n],
\end{equation}
which depends on both $f$ and $\phi$ and exhibits rich structures along different crystal directions (see Fig.~\ref{fig4}(c)).

The direction of the synthesized magnetic field, i.e., the sign of $\Phi_n$, determines the unidirectional transport along $b$-sublattice, {regardless of the strength $\kappa_n$ \cite{Cai2019}}. The chiral current flows to the right and left for $\Phi_n=\pi/2$ and $-\pi/2$, which results in asymmetric distribution of steady-state probabilities in the SL with respect to the probed site.  Such an asymmetry can be measured by comparing the  reflection signals of two probe fields \cite{Cai2019,He2022} in opposite directions and with opposite detunings. We define $R$ and $R'$ as the reflectivities of the probe field incident along $+x$ and $-x$ directions. $R(n\delta_2)$ characterizes the transport from the site $|b_{1,0}\rangle$ to  $|b_{-1,-2n}\rangle$ while $R'(-n\delta_2)$ describes the transport  from the site $|b_{-1,0}\rangle$ to  $|b_{1,2n}\rangle$ \cite{SM}. The chiral current along the crystal direction $[1,n]$ are measured by $\delta R_n\equiv R(n\delta_2)-R'(-n\delta_2)$ (see Fig.~\ref{fig4}(d)), which agrees with the phase diagrams in Fig.~\ref{fig4}(c). 

In conclusion, we demonstrate that the SL provides a highly tunable platform to study the Floquet dynamics in thermal atoms. We simultaneously observe dynamic localization and delocalization from a single transport spectrum. We also experimentally measure the phase diagrams of chiral edge currents in a (1+1)D lattice with both artificial magnetic fluxes and electric fields. {We can increase hopping strengths between TDSs to overwhelm the linear potentials, such that the motion of excitations is governed by the underlying 2D Bloch bands with the effective electric force being a perturbation. In such a regime topological non-equilibrium phenomena such as quantized energy pumping \cite{Martin2017,Kolodrubetz2018,Boyers2020,Long2021,Malz2021,Nathan2021} and space-time crystals \cite{Morimoto2017,Xu2018,Peng2019,Gao2021,Peng2022} can be investigated. Furthermore, we can introduce interaction in SLs by using additional lasers to couple atoms to the Rydberg states \cite{Li2020} to realize many-body phenomena beyond the short-range interaction \cite{SM}, such as the interplay between long-range nonlocal interaction and dynamic localization-delocalization transition \cite{Cao2022,Toh2022}.}

We thank Luqi Yuan for insightful suggestions. This work was supported by the National Natural Science Foundation of China (Grants No. U21A20437,  No.~11874322 and No.~11934011), the National Key Research and Development Program of China (Grants No.~2019YFA0308100, No.~2018YFA0307200 and No.~2017YFA0304202), Zhejiang Province Key Research and Development Program (Grant No.~2020C01019), Innovation Program for Quantum Science and Technology (Grant No.~2021ZD0303200), the Strategic Priority Research Program of Chinese Academy of Sciences (Grant No.~XDB28000000), and the Fundamental Research Funds for the Central Universities.

$\ $

$^{*}$These authors contributed equally to this work.

$^{\dagger}$hancai@zju.edu.cn

$^{\ddagger}$dwwang@zju.edu.cn

\end{document}